# CRITICAL FACTORS AND ENABLERS OF FOOD QUALITY AND SAFETY COMPLIANCE RISK MANAGEMENT IN THE VIETNAMESE SEAFOOD SUPPLY CHAIN

Thi Huong Tran

School of Economics and Management, Hanoi University of Science and Technology, Hanoi, Vietnam

## ABSTRACT

*Recently, along with the emergence of food scandals, food supply chains have to face with ever-increasing pressure from compliance with food quality and safety regulations and standards. This paper aims to explore critical factors of compliance risk in food supply chain with an illustrated case in Vietnamese seafood industry. To this end, this study takes advantage of both primary and secondary data sources through a comprehensive literature research of industrial and scientific papers, combined with expert interview. Findings showed that there are three main critical factor groups influencing on compliance risk including challenges originating from Vietnamese food supply chain itself, characteristics of regulation and standards, and business environment. Furthermore, author proposed enablers to eliminate compliance risks to food supply chain managers as well as recommendations to government and other influencers and supporters.*

## 1. INTRODUCTION

In recent years, global food consumers have witnessed a plethora of food quality and safety scandals, such as horsemeat in Europe (2013), the E. Coli Outbreak in the US (2015), Caraga candy poisonings in Philippines (2015), the Mars Chocolates contamination incident (2016), Punjab sweet poisoning in Pakistan (2016), Brazilian meat scandal (2017), and French salmonella baby milk (Lactalis) scandal (2018). The response of governments to these and earlier food safety problems has been to implement new food safety legislation or to release improved guidelines to food handling firms [1]. In fact, there are not just public regulations and standards but private standards that are issued by third party organization or group of retailers. Food quality and safety regulations aim to (i) protect the consumer's health, (ii) enhance economic viability, (iii) harmonize welfare and (iv) ensure fair trade within and between nations on foods [2].

However, according to Bolwig, et al. [3] and Gold, et al. [4], standards and regulations are international trade barriers especially for small and medium size agri-businesses in emerging and developing countries due to their difficulties in finance (access to capital), infrastructure (especially information technology, inspection and measurement equipment), and people (awareness, attitude, skills, and knowledge) [5]. In order to get a ticket to access the global market, compliance with food safety regulations of the home country and imported countries is the only sustainable method for all food chain members. In addition, risk based compliance management is demonstrated as an effective and efficient way to ensure food safety and quality in the food supply chain [6-10]. Aligning with the significance of compliance risk (to food quality and safety regulations), this paper will focus on analyzing critical factors of compliance risk and proposing enablers to ensure the successful management of this risk using secondary research combined with the expert interview method within the context of Vietnamese seafood supply chain.

The remainder of this paper is structured as follows. Section 2 presents a research background, which is based on comprehensive literature review about compliance risk in food chains and an overview about the Vietnamese seafood supply chain as well as compliance issue in this chain, to make a sound basis for raising research questions. In section 3, author introduces methodology used to conduct this research. Section 4 and 5 analyze and discuss findings about critical factors and enablers in compliance risk management in the case of the Vietnamese seafood supply chain. Section 6 concludes this paper with summary of findings and suggestions for future research.

## 2. RESEARCH BACKGROUND

This section introduces compliance risk in general and in the food supply chain in particular (theoretical background), and then review this issue in the case of the Vietnamese seafood supply chain (practical background). After that, the research problem and research question will be clearly stated.

### 2.1. Compliance risk in food supply chain

Over the past few decades, food safety objectives have become "an integral part of food chain management" [11]. In the same vein, food quality and safety management systems are generally becoming more stringent in different countries including developed and developing countries with a proliferation of regulations and standards [5, 12]. According to Marucheck, et al. [13], regulations are established by government agencies and it is compulsory to comply with them due to a sanction system while standards that establish *"norms or codified requirements for a product, such as material specifications or technical standards for performance"*, are *"developed by regulatory agencies, public organizations or industry associations"* without a sanction system. However, these two terms are often used interchangeably.

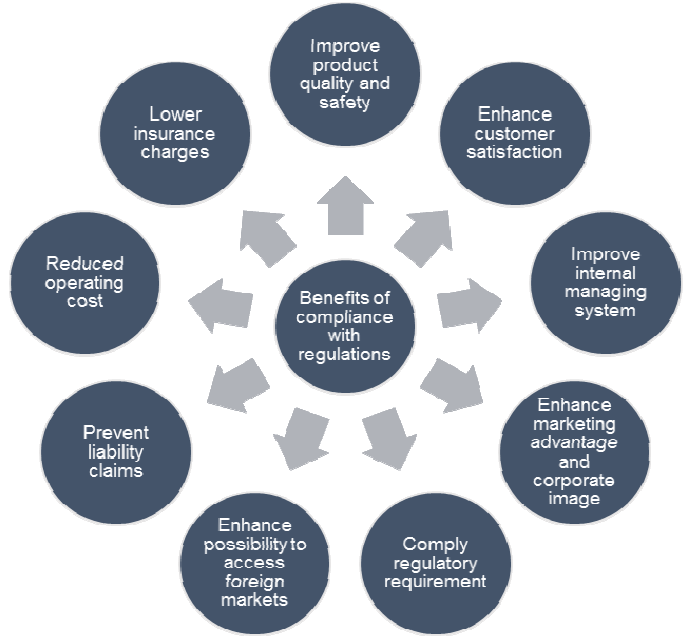

Figure 1. Benefit of compliance with regulations [5]

While some firms are against regulations and standards, most of business organizations recognize the great benefits from complying with regulations and standards. According to Marucheck, et al. [13], compliance can promote better product and service quality as well as giving a signal for a good management system. Mensah and Julien [5] synthesized a number of drivers and benefits of compliance with food quality and safety regulations as presented in figure 1.

Beside the above benefits, when implementing and maintaining compliance of food quality and safety regulations, food supply chain members also have to face many challenges such as: (i) internal supply chain factors ( high implementing cost, lack of trust, lack of money, lack of technology, lack of awareness, collaborative attitude, knowledge, skill of employees, and lack of collaboration of suppliers), (ii) regulation related challenges (rapid changes, friction, conflict, overlapping),  and (iii) external challenges (weak blame culture, lack of government support, and lax national inspection and sanctioning system) [5, 14, 15]. Indeed, the proliferation of standards and the bureaucratic burden can make it difficult for food business actors to comply [13, 16]. Furthermore, according to findings by  Zorn, et al. [17] and Zanoli, et al. [18], in some situations, the fact that compliance is costly, can stimulates opportunistic individuals to accept compliance risk instead of paying higher production cost and reducing their (short-term) profit.

As a result, compliance risk, or non-compliance risk which refers to the probability of a product or service being unable to meet the standards or requirements, is very common across food sectors as well as for actors in food supply chains and links to operational risk, cost risk  and reputation risk of the "farm to fork" chain [19].

Recently, food globally has been sourced from an increasing number of countries and regions, in some of which, non-compliance with the food law is more profitable for food business operators than compliance, especially considering the generally low probabilities of being inspected and detected [20]. For instance, evasion is preferable than compliance in many small scale businesses in developing countries [14]. That makes compliance risk management extremely complex in the context of even-longer and more international supply chains.

## 2.2. Enablers of compliance risk management in food supply chain

The risk management enabler was defined as the factors that "*enable risk management strategy implementation*" [21] and has its interchangeable terms as moderators, antecedents, and principles [22] These enablers can be autonomous, dependent, linkage, and independent in relation with other enablers [23] There are quite a few studies investigating challenges, barriers, and benefits of compliance, but only paucity of research on critical factors and enablers to successful implementation of a food quality and safety compliance system as well as compliance risk management framework in proactive and reactive approaches.

According to Jonker, et al. [14], the food safety and quality standards compliance system depends on two groups, namely company-specific factors and country-specific factors. Firstly, company-specific factors refer to capital restrictions (to implement and maintain new compliance system and to equip inspection machines), the maturity of the management system, and the attitude of managers and employees. Secondly, country-specific factors relate to current sanitation regulations, inspection capacity, assistance in information and finance of government. Mensah and Julien [5] conducted an empirical research on the successful implementation factors regarding compliance in the case of the UK's food supply chains. Their findings showed that there are four factor groups in order of increasing significance level as follows:

    i.  the first group includes four factors: government intervention, reward and recognition system for employees, linkages with learning- training centers, and measurement of employee satisfaction.

    ii.  the second group also includes four factors: awareness of employees about the importance of food quality and safety to the development of organization, usage of standard based operational procedures, continuous improvement, employee involvement.

    iii.  the third group comprises of education, training and supplier management,

    iv.  the fourth and the most important is top management involvement and commitment

## 2.3. The Vietnamese seafood supply chain

According to the Food and Agriculture Organization of the United Nations [24], more than half of seafood exports by value originate in developing countries such as China, India, Thailand, and Vietnam. Vietnam has been among the top ten exporters of fish and fishery products all over the world since 2004 and has become the third major exporter in the world seafood trading market since 2014 [24]. Having over 3000km of coastline and a dense river system, Vietnam possesses ideal conditions to develop a strong seafood industry, including both fisheries and aquaculture. The industry has developed dramatically in the past twenty years, created more than four million jobs in the Vietnamese labor force, contributed 4-5 % to GDP, and making up 5-6 % of total national turn-over. Vietnamese seafood products are consumed in 164 markets in 5 continents and have a great potential to develop in future [25]. The US, EU and Japan are the top three export markets of the Vietnamese seafood product with the total export percentages (by value) in 2015 being respectively 20%, 18%, and 16%. Figure 2 and 3 show the total production and export value of the Vietnamese seafood (including fisheries and aquaculture products) in recent years.

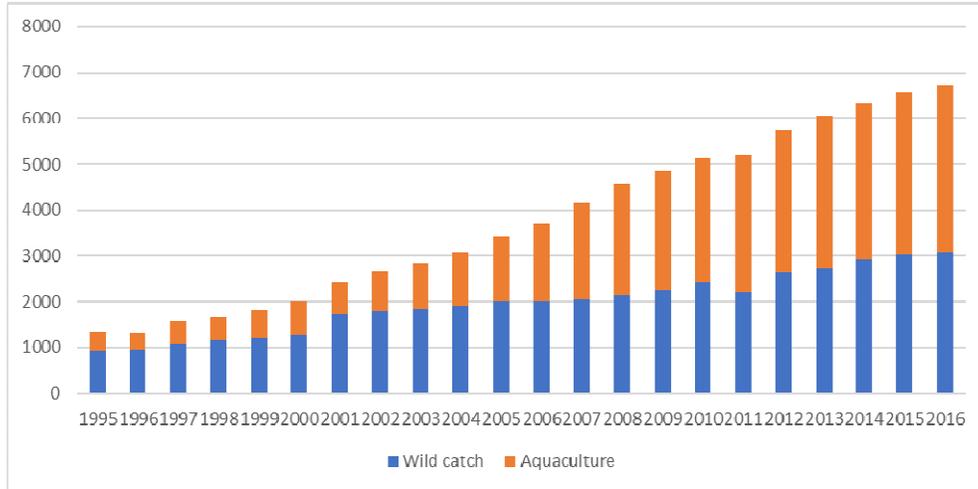

Figure 2. The yield of Vietnamese seafood from 1995 to 2016 (thousand tons). Source: [26]

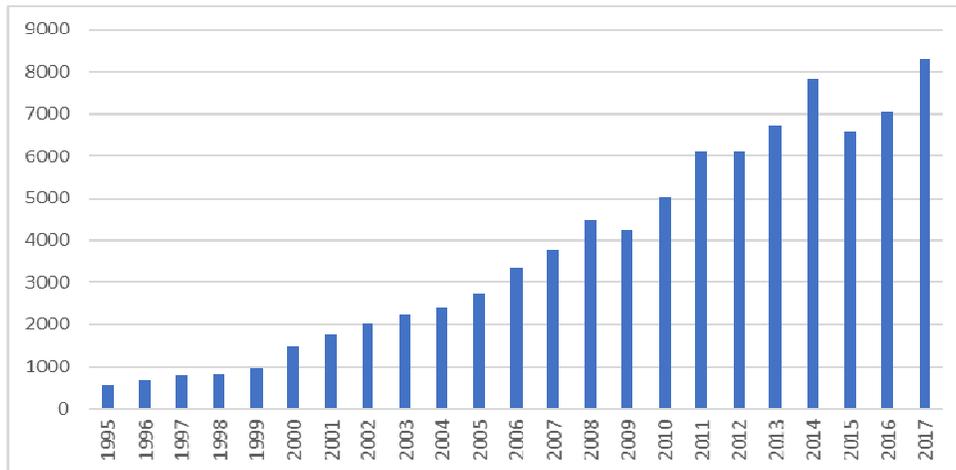

Figure 3. Vietnamese seafood export from 1995 to 2017 (US$ million)
Source: [26, 27]

According to Vietnam Directorate of Fisheries- FISTENET [28] and Vietnam Associate of Seafood exporters and producers - VASEP [26], the four most important products of the Vietnamese seafood industry are shrimp (accounts for 44% total seafood export value in 2015), pangasius (23%), tuna (7%) and molluscs (8%) (including clams, oysters, mussels, squid and cuttlefish).

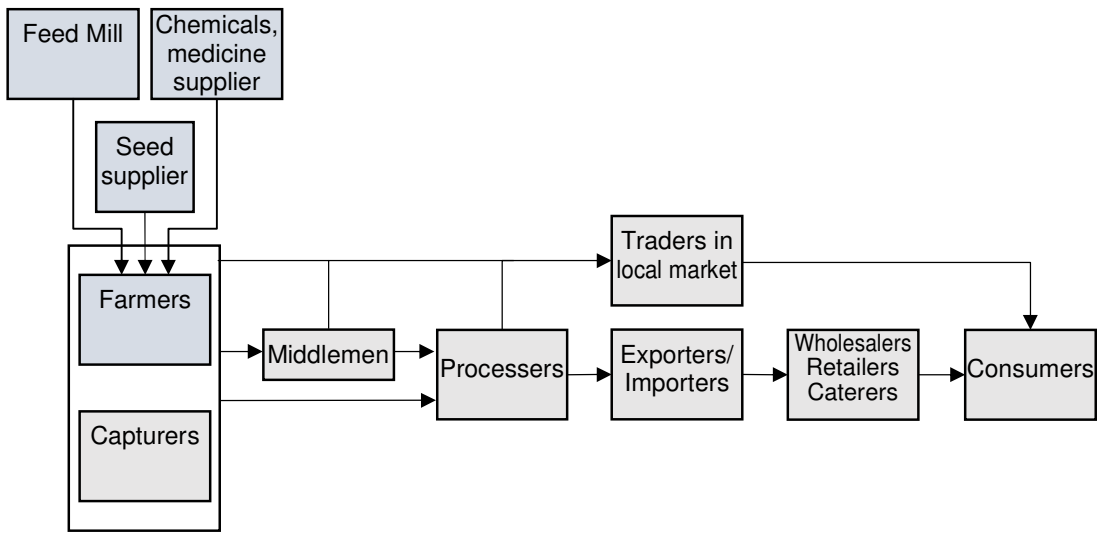

Figure 4. Structure of the Vietnamese seafood supply chain [26]

Figure 4 describes the general structure of the Vietnamese seafood supply chain. There are a number of main actors in this chain, including (i) input suppliers (for aquaculture seafood supply chain, including hatchery nurse, feed mill, medicine and chemicals suppliers); (ii) capturers (fisheries chain) or farmers (aquaculture seafood supply chain) who buy hatcheries, feed them and use supplementary chemicals and medicine if necessary; (iii) middlemen who gather seafood (material) for traders in local markets, processors or exporters; (iv) processors who add value and lengthen the life cycle of seafood product; (v) exporter/importer to global markets or traders in local market; (vi) wholesalers, retailers and caterers who distribute seafood products to consumers; and (vii) consumers who ultimately decide the survival and development of seafood supply chains.

In recent years, the export value of Vietnamese seafood has nosedived due to a number of bottlenecks and external challenges. Suzuki and Vu [29] analyzed a number of reasons leading to these bottlenecks in the Vietnamese seafood supply chain, including (i) the decrease of marine stock due to unsustainable fishing methods for many years, (ii) difficulties in catching business due to weather conditions, fuel price, labor and capital cost etc., (iii) deficiency in quality of broodstocks, (iv) the continuous increase of imported feed, (v) outbreak of diseases, (vi) the improper usage of pesticides, antibiotics, and chemicals, (vii) the inappropriate master plan for aquaculture, (viii) a lack of knowledge, capital, and coordination, (ix) non-tariff barriers from imported markets, (x) increasing complexity of food quality and safety standards.

Besides that, according to Shieh-Liang, et al. [30], poor supply chain management, especially the thinking and viewpoints of supply chain managers is also a significant reason contributing to the reduction in export value of Vietnamese seafood.

VASEP [31] defined seven major challenges in the Vietnamese seafood industry in 2017, including (i) climate change, drought and salinization in Mekong Delta river areas which have decreased seafood farming acreage and yield significantly; (ii) the increase of technical barriers and trade protection of importing markets; (iii) the high production cost in comparison with Thailand and India; (iv) a shortage of raw material for processing and exporting; (v) intense competition in terms of productiveness, quality, cost, marketing and trade promotion; (vi) reputation risk due to negative communicated information; and (vii) shortcomings in regulation and administrative procedures.

## 2.4. Compliance risk in the Vietnamese seafood supply chain

Aligning with the fact that Vietnamese seafood products have been imported into 164 markets in 5 continents [26], there are a large number of food quality and safety regulations and standards which the Vietnamese seafood industry has to comply. *Firstly,* the Vietnam national regulations system on food safety and quality comprises food safety law No. 51/2001/QH10 and a number of national technical regulations e.g. QCVN 02-10: 2009/BNNPTNT ("Aquatic product purchases - Food hygiene and safety assurance condition"). *Secondly,* the Vietnamese seafood supply chain must comply international regulations established by the FAO, WHO, WTO (Agreement on the Application of Sanitary and Phytosanitary Measures (SPS) and Agreement on Technical Barriers to Trade), and FAO/WHO-Codex Alimentarius Commission (code of practice for fish and fishery products - CAC/RCP 52-2003). *Thirdly,* required regulations of specific importing markets i.e. the United States, the European Union, Japan, Korea, Russia, Australia, New Zealand, Taiwan, Ukraine, Brazil, … For example, the Food Safety Modernization Act by the U.S Food and Drug Administration (FDA), Maximum Residue Limits (MRLs) of chemicals contaminated in food products; Regulation EC No 1333/2008 on food additives; Regulation EC No 852/2004 on the hygiene of foodstuffs; Regulation EC No 882/2004 on "official controls performed to ensure the verification of compliance with feed and food law, animal health and animal welfare rules". *Fourthly,* there are required standards from specific importing markets such as ISO 9000 (Quality Management System), ISO 22000 (Food safety management system), HACCP (Hazard Analysis and Critical Control Point), GMP (Good Manufacturing Practice), BAP (Best Aquaculture Practices), GAP (Good Aquaculture Practices), SQF (Safe Quality Food) 1000 and 2000, BRC (British Retail Consortium), IFS (International Food Standard), and HALAL (Standards for food under the Islamic Law).

Food quality and safety regulations and standards have been and still are the significant barriers to the export of Vietnamese seafood products. According to Vietnam National Agro-Forestry-Fisheries Quality Assurance Department (NAFIQAD), the total number of food safety alerts from the EU about Vietnamese seafood in the first ten months of 2016 increased by 2,2 times in comparison with the entire year 2015. As a result, there are an increasing number of returns and exporting restrictions. Recently, 11 shipments of Vietnamese seafood have been turned back from the EU due to heavy metal residue exceeding allowed limit. According to UNIDO consensus, Vietnam is at the top of countries having seafood products restricted to import to the US and EU. In the same vein, there are a number of other countries that have banned imports of certain seafood products coming from Vietnam in a given period due to food quality and safety problems.

According to Bridonneau [32], complying with food quality and safety standards is the most significant non-tariff barrier in the agricultural sector, and is perhaps the greatest obstacle of exporting to middle and high-end markets for Vietnamese food products, especially seafood products.

As a result, "what are the critical factors of food quality and safety compliance risk (abbreviate as compliance risk) and enablers to compliance risk management?" is the urgent question for practitioners in the Vietnamese seafood supply chain as well as the research question which this paper is raising to find out answers. To this end, research framework was designed and is presented in the next section.

## 3. RESEARCH DESIGN AND METHODOLOGY

To investigate compliance risk in the Vietnamese seafood supply chain, researchers utilized an exploratory research strategy through a three-step approach. At the first stage, a review of research papers as well as industrial reports on compliance risk in the food supply chain and current status of compliance risk in the Vietnamese seafood supply chain was conducted to gain a comprehensive understanding of the research problem and to identify critical factors from the point of view of all members and stakeholders. Next, semi-structured expert interviews were carried out to define critical risk factors and solutions to deal with the food quality and safety compliance risk from the point of view of processors and exporters who have thus far not received much attention in literature. In the final stage, evidence from literature, expert interview and industrial information were analyzed and synthesized to find out nature of compliance risk in the Vietnamese seafood supply chain as well as recommend risk management solutions to practitioners.

The expert interview has been a popular qualitative method in exploratory research strategies in social science [33, 34]. The semi-structured expert interview is a fairly open research framework which uses several key questions to define the major areas to be explored and also allows experts to show their knowledge of relevant issues in a two-way communication (Silverman, 2000). In this research, author interviewed managers of quality assurance departments and directors of four seafood processing and exporting companies in 2016, with major questions concerning their perceptions about compliance risk, their difficulties in complying with food safety and quality regulations, and the causes and solutions to deal with this issue in their cases.

Our critical factor analysis and proposed solutions/enablers to compliance risk management in the Vietnamese seafood supply chain are based on suggestion from interviewees, supply chain structure approach, and the framework of "the table of eleven". "The table of eleven", which was developed by Dr. Ruimschotel and the Law Enforcement Expertise Centre, Dutch Ministry of Justice (LEEC) and is widely applied in agri-food supply chain management, comprises eleven factors to encourage and enforce compliance [9, 35].

# 4. CRITICAL COMPLIANCE RISK FACTORS IN THE VIETNAMESE SEAFOOD SUPPLY CHAIN

Through interviewees, literature and industrial reports, findings about critical risk factors which themselves will be the identifiers for enablers of compliance risk management in the Vietnamese seafood supply chain are investigated and presented in figure 5. Critical compliance risk factors are divided into three main groups: (i) the main actors in the Vietnamese seafood supply chain (net-chain members), (ii) regulation and standards, and (iii) business environment e.g. government and other stakeholders (consumer, industrial association and communities).

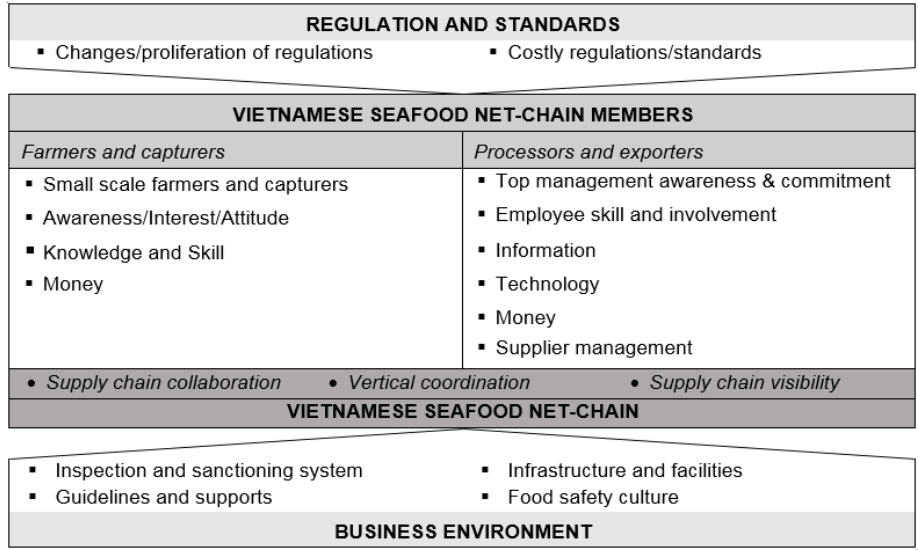

Figure 5. Critical compliance risk factors in the Vietnamese seafood supply chain

## 4.1. The Vietnamese seafood net-chain members

The most important factors originate from the inside of the Vietnamese seafood supply chain itself. There are two main actors which have critical impact on compliance system and performance as follows.

### 4.1.1. Farmers and capturers

▪ Small scale farmers and capturers
Farmers and capturers are the first major actors in the Vietnamese seafood supply chain. They are mostly undertaking business at the micro to small and medium scale with limited human resource and funds (for example, according to DOF [36], 80% of shrimp production in Vietnam is in small scale household farms). That leads to other risk factors and bottlenecks in food quality and safety compliance for the whole supply chain.

▪ Awareness, interest, and attitude
One of the most significant factors in compliance is the awareness of the necessity and importance of complying [37]. Most of farmers and capturers perceived regulations and standards as complexities and difficulties, not a beneficial market tool [37]. Findings from the recent survey of Hansen and Trifković [38] showed that according to fish farmers, compliance only benefits directly the upper and a part of middle class farmers instead of lower and some middle class farmers [39]. According to Trifković [37], only 20% of household owned farms apply standards in the case of pangasius sector, they perceived that the antibiotic residues index is the only one requirement for food quality and safety [40]. Even when benefits were recognized, due to some prohibiting factors such as pursuing low-end markets, short term profit aims, and facing financial constraints, farmers and capturers have very low interest in compliance, resist changes in farming and capturing procedures, and feel risk aversion when applying new systems.

▪ Knowledge and Skill
Most Vietnamese farmers and fishermen have developed their skills in farming and fishing through experience rather than formal education and training [41]. As a result, their knowledge does not satisfy new requirements of international regulations and standards, especially in solving healthcare problem, storage and transportation. According to VASEP, most Vietnamese seafood farmers have limited education and training level and develop spontaneously without long- term plan or with little planning. Especially, there is a lack of knowledge and experience in implementing, maintaining and controlling compliance with international regulations and standards.

- Money
  One significant risk factor as well as burden for overcoming all above and other bottlenecks is that of restricted access to capital by smallholder farmers in emerging and developing countries [3]. Most farmers in the Vietnamese seafood industry depend on credit from banks or financial/feed support from their customers (middlemen, processors or distributors). They lack money to afford cost of consultancy, cost of certification, cost of upgrading technology and cost of innovating their farming and fishing procedures.

### 4.1.2. Processors and exporters

- Top management awareness and commitment
  As an interviewee stated, there are three groups of seafood processor, namely small scale, medium scale and large scale. Each of them has a different awareness and commitment with regulations compliance. While the large-scale processors and exporters always update to changes and new regulations/standards to satisfy their customers, medium and small-scale organizations only try to maintain the basic regulations/standards, and only slowly adapt to new ones, thus easy to get alerts and lose purchasing orders when consuming markets recognize new healthy risk and release new regulations.

- Employee skill and involvement
  Seafood processing is depicted as a sector with high turnover rate of employee, which challenge processors in maintaining a stable quality and safety management system. Another problem is the resistance to change of employees when adapting to a new regulation or standard. In addition, due to the fact that most of the processing facilities located near to farming areas or fishing stations which are not big cities in Vietnam, it is difficult to recruit a professional and high skilled staff.

- Information
  Information sources which support identifying and interpreting regulations to compliance, includes information from local authorities (leaflets, websites, training, inspection visits, advisory visits), industrial and trade associations (newsletter, leaflets, websites, individual advice), and other sources such as visits, training, briefing notes, reports from partners and consultants. The facility location issue, as stated above, and poor information system infrastructure lead to the difficulty in dissemination and the lack of access to those information sources. That causes slowness and inappropriateness in updating and complying with food quality and safety regulations and standards.

- Technology
  The old-fashioned technology in processing, inspection, storage and transportation, especially in small and medium size processors and exporters, causes inconformity in input, in-process products, and finished goods, decreases food quality as well as harm food safety and restricts traceability.

- Money
  The majority of Vietnamese processors and exporters are at small and medium scale. After a rapid growth period, Vietnamese processors and exporters have faced the risk of bankruptcy due to lack of capital, financial management skill and increasing interest rate since 2012. That results in companies being unable to afford a food safety compliance system and then follow short term value (beside the reason that compliance is costly and results in higher production costs [17, 18]). As a result, compliance risk consequences take effect and entail many others operational and financial difficulties.

- Supplier management
  Quality and quantity of supplied seafood have a significant impact in food safety compliance. Due to lack of collaboration with suppliers and sourcing capacity, in some situations, Vietnamese seafood processors have had to purchase raw seafood which does not meet quality and safety requirements in order to maintain production and make a bet on avoiding compliance risk in high-end markets or to sell to low-end markets.

### 4.1.3. Overall seafood supply chain

From the perspective of the overall seafood supply chain, the weakness of supply chain collaboration, vertical coordination and supply chain visibility are the critical factors hindering Vietnamese seafood producers from meeting regulations and standards compliance. As interviewees in processing and exporting companies stated, although they are willing to comply food safety regulations and standards, small and household scale capturers/farmers do not want or cannot afford to apply for and update regulations and standards. Besides that, traceability, which is "a precondition of demonstrating compliance" [42], is not developed completely in the Vietnamese seafood supply chain. One significant reason is that "the smaller scale farmers are, the more difficult and costly traceability becomes" [42].

Middlemen play an important role in the Vietnamese seafood supply chain because most farmers and capturers do not have business management knowledge, especially bargaining power and skills, and there is a lack of synchronous coordination of processors, exporters and governments. Middlemen supply inputs in farming, harvesting for farmers/capturers and work as collectors to collect small orders from small household farmers/capturers for processors and exporters. According to Kusumawati, et al. [43], middlemen are the key member, however, there are a lot of opportunistic middlemen who encourage breaking rule behaviors i.e. using banned/unknown origin chemicals and medicines, cover up safety regulation violations, do not keep traceability records etc. Contract farming and vertical coordination are highly appreciated solutions for this issue, but they have not been successful due to conditions in contracts that have not been respected by all related parties.

## 4.2. Regulations and standards

Regulations and standards for food safety are mostly imposed by governments or third-party organizations in importing countries (mostly, high-end consumer markets i.e. Japan, the European Union, and the United States). Although most regulations were made to bring benefits for all related parties from consumers to producers, in many situations, food quality and safety regulations and standards cause a number of difficulties and potential compliance risks to food businesses.

### 4.2.1. Changes/proliferation of regulations and standards

According to Buzby and Mitchell [44], differences in food quality and safety regulations and standards among importing and exporting countries without appropriate harmonization can cause friction and impede international food trade.

Valdimarsson [45] stated that the usage of private and voluntary standards from a retailer is a way to enforce compliance. However, there are too many different and overlapped standards from different import markets. Rapid changes and the proliferation of regulations and standards due to the raising of potential hazards and risks, cause confusion for many stakeholders [46] i.e. disordering smooth operations of seafood businesses due to a requirement to continuously update changing information, altering their process from sourcing, producing, to labelling, etc.

### 4.2.2. Costly regulations and standards

For a developing economy, complying with food quality and safety regulations and standards of importing markets (majority is developed countries) is not only a difficult task, but also an expensive and long-term plan. The case of implementing VietGAP to shrimp farms is an example, although the Vietnamese government pay administrative, assessment, and training costs for farmers, there are still many types of cost left over [39] i.e. cost to improve capacity of farm (i.e., digging, cost to improve water quality and waste management, etc. Costly regulation together with difficulties in capital entail non-compliance behavior of several actors in the Vietnamese seafood supply chain, especially, in the context of an incomplete traceability system where compliance risk is harder to control.

## 4.3. Business environment

There are four considerable factors leading to compliance risk in the Vietnamese seafood supply chain as follows.

### 4.3.1 Inspection and sanctioning system in the Vietnamese seafood supply chain

The government plays an important role in enforcement (through inspection and the sanctioning system) and in fostering voluntary compliance (through training, guidelines and supports). Indeed, according to Fairman and Yapp [47], the enforcement together with strict inspection and sanctioning system are external motivators of compliance with food safety legislation in small and micro-businesses.

According to Raakjær Nielsen [48], compliance behavior depends on *"the economic gain of breaking regulations compared to the risk of being detected and the economic sanctions of rule breaking"*. In the same vein, Van Asselt, et al. [9] showed that there are six enforcement dimensions enforcing compliance behavior, including (i) probability of third party reporting, (ii) probability of inspection by authorities, (iii) probability of detection, (iv) selectivity of inspection, (v) probability of sanction, and (vi) severity of sanction. In many situations, non-compliance with the food regulation is more profitable than compliance, especially when doing business in the context of low probabilities of being inspected and detected [20].

In the context of the Vietnamese seafood supply chain, according to Bridonneau [32], there is a lack of both encouragement and enforcement from the authorities governing compliance. Firstly, the specialized inspection human resource for food safety inspection and sanctioning system in Vietnam is limited (less than 1,000 people) in comparison with other countries such as Thailand (more than 5000 inspectors only in Bangkok) and Japan (12000 food safety inspectors) [49]. That fact entails a low frequency and probability of inspection. As a result, there are various gaps in controlling the violation of food quality and safety regulations, which leads to unfair competition and decreased incentives for developing and maintaining a food safety system [49]. Secondly, the reliability of inspection/testing laboratories is quite low due to selectivity procedure, analytical capacity, and the quality of inspecting machines [50], that caused a lot of consignments had passed national testing but failed at the testing stage of the importing markets.

In addition, according to Mr. Nguyen Huu Dzung, former vice chairman of VASEP, government agencies such as NAFIQAD only set stringent control at the final stage (processing), while earlier stages are much more critical in compliance risk management [51]. Besides that, the unclear responsibilities between various departments in MARD may lead to ineffective control of food safety in the Vietnamese seafood supply chain [52].

### 4.3.2. Guidelines and supports from government and industrial associations

There are quite a few multi-layer governments, NGO and industrial associations which provide guidelines, technical and financial supports for the main actors in the Vietnamese seafood supply chain [29], such as MARD, NAFIQUAD, VASEP, VINAFIS, CBI, etc. However, there is still a lack of clear and synchronous guidelines. Financial supports still lacking with slow implementation progress together with many restrictions for small scale farmers and fishermen who dominate operations in the Vietnamese seafood chain.

### 4.3.3. Infrastructure and facilities

According to Luu, et al. [50], the infrastructure and facilities, such as fishing ports and fish markets, are quite poor, and that has affected the potential non-compliance in food quality and safety regulations and standards. A survey of Tran, et al. [53] show that the majority of shrimp farmers operate in rural areas without roads, and that this leads to the existence of various levels of middlemen who mostly are uneducated, unaware and even ignore of the importance of food safety issues [54] and may distort information about food safety regulation as well as impacting on the traceability system. The development of information communication infrastructure in the Vietnamese agricultural sector and seafood supply chain is still in its infancy [55], and this leads to the lack of traceability to take control of food safety compliance.

### 4.3.4. Food safety culture

Public concerns about food safety contribute to place pressure on government agencies to be more strict in controlling compliance and on business actors to be more proactive and responsible in complying with food safety regulations and standards [56]. However, due to the characteristics of a developing economy, concerns, blame culture and power of consumers in Vietnam about food safety is just in its infancy. That is the critical cause of low food safety culture and entails difficulties in ensuring compliance across the whole Vietnamese seafood supply chain when pursuing export-oriented development.

## 5. ENABLERS TO COMPLIANCE RISK MANAGEMENT IN THE VIETNAMESE SEAFOOD SUPPLY CHAIN

While food quality and safety regulations and standards are essential, the emerging questions are what can provide effective incentives to comply with regulations and standards [13], what can support the compliance process and eliminate compliance risk in the supply chain, and what is the appropriate compliance risk management framework.

### 5.1. Enablers from influencers of the Vietnamese seafood supply chain

Enablers form influencers and supporters of the Vietnamese seafood supply chain can be divided into two main groups. While the first group relates to the supports which can encourage compliance behaviors, the second group refers to the inspection and sanctioning system which create enforcement power to eliminate non-compliance attitude and activities.

### 5.1.1. Supports from government and industrial associations

There are a number of measures which the government can use to foster safety compliance in the Vietnamese seafood supply chain, such as (i) developing guidelines to implement formal food safety management systems, (ii) providing training courses and workshops to enhance knowledge, skills and updating as well as harmonizing new regulations and standards, (iii) supplying technical and financial support in capturing, farming and processing, (iv) improving infrastructure and facilities; and (v) improving communication between business actors and governments [41, 50].

Guidelines, education, training, and workshops by government and industrial associations provide valuable technical support to improve understanding, knowledge, awareness and skill in complying and updating food safety regulations and standards for the main actors in the Vietnamese seafood supply chain, especially small-scale farmers, capturers and processors [50, 57, 58].

Financial support, a masterplan for the whole seafood industry i.e. land consolidation, cooperation, and grouping micro/small scale household farmers/capturers, and contract farming will help them get access to credits easier and reduce compliance costs (consultant fee, implementing, audit and certificating fee).

### 5.1.2. Strict inspection and sanctioning system

While guidelines and support from influencers encourage spontaneous compliance dimensions, the inspection and sanctioning system are enforcement dimensions that have a direct effect on compliance behavior through the increased probability of reporting/inspection and detection as well as the severity of non-compliance sanction.

As there is a current laxness in the inspection and sanctioning system, the stricter control of middlemen and farmers as well as traders in using and distributing unknown origin/banned chemicals, antibiotics and pesticides is a significant enabler of compliance risk management. This action will also contribute to improve food safety culture in business actors and domestic consumers.

The effectiveness of the inspection and sanctioning system can be improved through increasing quantity and quality of inspectors and laboratories as well as the frequency of inspecting, modernizing inspection machines, and enforce a traceability system throughout supply chains.

## 5.2. Enablers from main actors in the Vietnamese seafood supply chain

The enablers to manage compliance risk should be originated from the supply chain itself, which includes a compliance risk management framework, information sharing, traceability system and vertical coordination practices.

### 5.2.1. Compliance (risk) management framework

The first enabler from the Vietnamese seafood supply chain can be a compliance risk management framework/procedure with a clear code of conduct and responsibility for each person, position, department in an organization and each member in the supply chain to comply food quality and safety regulations. Developing, communicating and maintaining this framework can simultaneously encourage voluntary and enforce compulsory compliance in the supply chain.

This research proposed an example food safety compliance risk management process (figure 6) that comprises six main stages. At the first stage, top managers should understand the importance of and commit to comply with food safety regulations and standards. Next, every related food safety regulation and standard needs to be registered and periodically updated. In the implementation process, an iterative three step compliance risk management will be done to establish a rigorous compliance action plan with all potential enablers and proactive as well as reactive responses. In addition, all activities in the compliance management framework will be reported and openly communicated throughout the organization/supply chain in order to involve everyone in continuous improvement and eliminate potential compliance risks. The responsibility of person in charge should be clearly defined for each stage and action when implementing this risk-based compliance management framework.

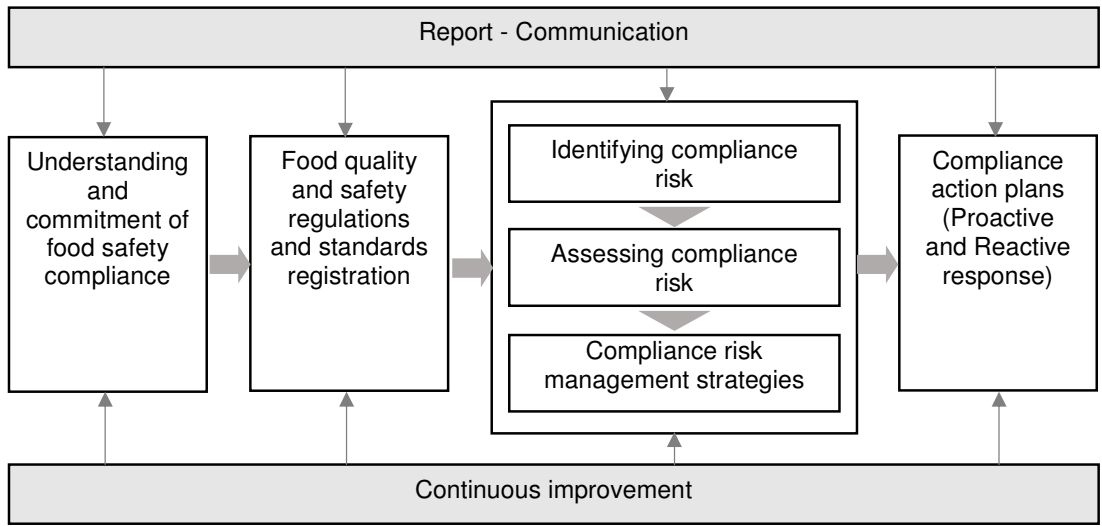

Figure 6. Food safety compliance (risk) management process adapted from Australian Standard for Compliance Programs (AS3806-2006)

### 5.2.2. Information sharing

According to Arpanutud, et al. [60], the more food safety information is received, the higher likelihood there is to adopt food safety management systems. In addition, information sharing between processors and farmers/capturers, which is an antecedent of supply chain visibility and encouraged by the formation of long-term relationships and trust, will limit unethical and opportunistic behaviors, especially of middlemen [61, p.106]. In the case of the Vietnamese seafood supply chain, information, which needs to be shared to eliminate compliance risks, comprises requirements, regulations, and standards about food safety in target markets, immediately and long-term consequences of non-compliance behaviors, guidelines, experience and managerial as well as technical knowledge in implementing compliance management systems, etc.

### 5.2.3. Supply chain traceability

In the context of increasingly complex supply and distribution chains, traceability has become especially important. An effective traceability system in the food chain can be measured through breath of traceability (number of variables/attributes are tracked and traced), depth of traceability (how far upstream and downstream of the food chain can be traced), and precision of traceability [61, p.154]. There are six elements needed to trace in food supply chain including (i) product traceability (location of products at every stage), (ii) process traceability (type and sequences of activities influencing on product), (iii) genetic traceability (genetic constitution of the product), (iv) input traceability (type and origin of inputs i.e. chemical sprays, feed, additives), (v) disease and pest traceability (traces the epidemiology of pests and emerging pathogens that may contaminate food), and (vi) measurement traceability (quality of measurements relating to individual measurement results to accepted reference standards) [62, 63]. A robust traceability system facilitates the prevention of food fraud, or non-certified food products being passed off as certified product [46]. To establish a traceability system, all members in the Vietnamese seafood supply chain from hatchery farms, raising farms, capturers, middlemen to processors and exporters have to record documents about name/codes, location, quality and safety certification, etc. of all inputs, indigents, processes and outputs (one step backward and one step forward).

### 5.2.4. Vertical coordination

In the Vietnamese seafood supply chain, the asymmetric information between buyers and suppliers has generated a lot of problems including fluctuation of input price, lack of quantity and quality of material, and the ineffectiveness of compliance management system. As a result, vertical coordination (typically vertical integration and contract production) which has become the emerging trend in food supply chain in both developing and developed countries [37, 64], is a good choice. Vertical coordination can help both small and medium scale farmers and processors to have a greater chance to get access to credits from governments and NGOs.

In addition, contract production and full vertical integration allow contracted farmers to receive support from downstream buyers such as training, finance, input supply or technical and managerial advice [65] and to take these farmers under closer monitoring of the processors [29]. As a result, the traceability and compliance system can be ensured with more certainty through controlling amount and origin of fingerlings, feed, the type of drugs and chemicals used, labeling to track and trace, maintaining stable quality and quantity of raw material for processing, etc. That allows stringent quality and safety regulations and standards from governments and imported markets to be satisfied more easily.

In future, it is also essential for the Vietnamese seafood industry to vertically integrate to upstream traders, such as by directly doing business with retailers in imported markets to become more active in food safety regulations and standards registration instead of totally depending on importers as is the current status.

## 6. CONCLUSIONS

All stages of the seafood supply chain ranging from fishery/farming, processing, distributing, logistics and retailing are subject to food safety and quality regulations and standards in order to provide safe food for human consumption [61, p.174]. For many developing countries including Vietnam, meeting minimum SPS is already a major challenge [45]. There is the fact that most business actors in the Vietnamese seafood supply chain are small in scale, which should receive special consideration (as FAO Technical Guidelines for Aquaculture Certification), but food quality and safety should not be compromised to ensure global public health and sustainable development [39].

In this vein, author has considered critical risk factors in compliance with regulation and made key contributions in proposing enablers to eliminate compliance risks to food supply chain managers. As a saying that "risk never be zero but can be eliminated", one of the most important recommendations is to increase the awareness of risk in overall supply chain members through a compliance (risk) management framework as well as through information sharing together with guidelines and support from influencers. Equally important, risk-based compliance management systems might represent an effective tool to decrease the incidence of non-compliance. Coordination and traceability are essential in supply chain management; this situation is no exception. Last but not least, the role of government and industrial associations in encouraging and enforcing food safety regulation compliance is undeniable. In the future, further empirical confirmatory research is needed to exam criticality of risk factors and efficiency of recommended compliance risk management enablers. In addition, the risk in complying with sustainability related regulations and standards is an emerging research topic.

## ACKNOWLEDGEMENTS

The author would like to thank the three Vietnamese seafood processing and exporting companies who contributed significantly to my empirical research.

**Author**

Thi Huong Tran is a lecturer at School of Economics and Management, Hanoi University of Science and Technology, Vietnam. She received her doctoral degree at Institute for Transport and Logistics Management, Vienna University of Economics and Business, Austria. Her research interests are supply chain management, quality management, and supply chain risk management. 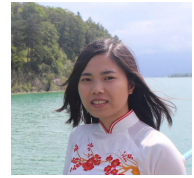